# Modified electrical transport probe design for standard magnetometer

B. A. Assaf,[1] T. Cardinal,[1] P. Wei,[2] F. Katmis,[2,3] J.S. Moodera[2] and D. Heiman[1]

[1]Department of Physics, Northeastern University, Boston, MA 02115
[2]Francis Bitter Magnet Lab, Massachusetts Institute of Technology, Cambridge, MA 02139
[3]Department of Physics, Massachusetts Institute of Technology, Cambridge, MA 02139

*Abstract*

Making electrical transport measurements on a material is often a time consuming process that involves testing a large number of samples. It is thus inconvenient to wire up and rewire samples on to a sample probe. We therefore present a method of modifying Quantum Design's MPMS SQUID magnetometer transport probe that simplifies the process of sample mounting. One of the difficulties to overcome is the small diameter of the sample space. A small socket is designed and mounted on the probe so that various samples mounted on individual headers can be readily exchanged in the socket. We also present some test results on the topological insulator $Bi_2Te_2Se$ using the modified probe.

## I. INTRODUCTION

The ability to make resistivity and Hall measurements efficiently at low temperatures and high magnetic fields is critical to research productivity. At present, various technological solutions are being sought to improve the measurement capability of materials that exhibit correlations between magnetism and electric transport. The anomalous Hall effect (AHE) in magnetic semiconductors[1] and metals[2], the quantum Hall effect (QHE) in graphene,[3] as well as various exotic Hall signatures in topological insulators are examples of such magnetoelectric phenomena that are making research headlines. Also, the fundamental understanding of quantum transport processes such as weak localization (WL) and quantum oscillations[4] requires similar experimental apparatus.

Quantum Design, Inc. provides a flexible solution with their SQUID MPMS® magnetometer, which can be equipped with an EverCool® dewar. The basic instrument allows one to measure magnetic properties to fields of B = 5 T or higher and down to temperatures of T = 1.8 K. Although the main function of the instrument is magnetization measurements, an optional electrical transport probe allows users to make use of the 5 T magnet in order to perform electrical conductivity measurements down to liquid helium temperatures.

The standard transport probe allows one to make transport measurements in a magnetic field. However, it does not offer the user the option of mounting and demounting samples very easily; each sample must be mounted and wired to the probe by hand. This process is time consuming. And in many cases one has to deal with the reality that samples often turn out to be defective or have the more common problem of defective wiring especially where the wires contact the sample. Such obstacles are notoriously known for slowing down the pace of research. In addition, in many cases one is simply required to investigate a large number of samples in a short amount of time. This cumbersome process of mounting and wiring samples can therefore significantly decrease productivity.

In this article we show how the optional Quantum Design transport probe can be modified in order to accelerate the process of sample mounting. A circular 10-pin header is modified and mounted at the bottom end of the probe. The mounted header is then used as a socket to support a second header on which a sample can be mounted. This way, various samples can be mounted and wired up on separate headers that can be interchangeably plugged into the socket. We demonstrate this design by carrying out transport experiments on a Hall bar patterned thin film of the topological insulator material $Bi_2Te_2Se$ at temperatures down to T = 1.8 K.

## II. INSTRUMENTATION DESIGN

The two ends of the optional electrical transport probe are shown in Fig. 1. The top end of the probe shown in Fig. 1(a) has a rectangular 10-pin electrical socket where a connector and multiwire cable can be attached. The bottom end of the probe shown in Fig. 1(b) has a row of 10 small screws for electrical connections that are internally factory-wired to the rectangular connector at the top of the probe. Below the 10 small screws at the bottom end there is a flat area (not presently visible in Fig. 1b) where samples can be attached. In normal operation samples are mounted to the flat area and wires must be applied to link the sample to the screw connections. The pictured probe has been modified by attaching a small circular 10-pin header to the end of the probe. A standard 10-pin "Augat Interconnect 8059-2G10" header is used, but the diameter must be machined down to a diameter of about 7.5 – 8.0 mm in order to provide sufficient clearance in the 9 mm inside diameter of the sample tube of the magnetometer. After machining, the header pins are nearly but not quite exposed on the periphery, as the diameter encircling the outside of the pins is 7.0 mm. Figure 2 shows the header before (a) and after (b) machining the outer diameter. The header is then mounted on the probe end by a small diameter (2-56) nylon screw placed through a hole in the center of the header. A small plastic bracket is fashioned to accept the screw holding the header and is mounted on the flat part of the probe end with two small screws. The socket pins are then wired to the row of screws on the probe. Insulated copper wires were soldered to the pins using lead-tin solder, which is diamagnetic and superconducting only below T = 7.2 K and 0.08 T,[5] and then attached to the contact screws. Once attached and wired, the socket is ready to accept plug-in headers that are mounted with samples. Many samples can now be interchanged by simply plugging and unplugging the headers into the socket.

Sample headers are fabricated by machining down standard 10-pin Augat headers to the same 7.5 – 8.0 mm diameter as the socket. Small pieces of sample were cleaved in order to fit conveniently on the headers. Samples were attached to the headers using common rubber cement. Rubber cement provides good bonding to temperatures as low as T = 2 K, yet remains somewhat elastic a low temperature and thus limits strain on the sample. Gold wires (about 0.1 mm diameter) were soldered to the header pins and the sample using indium, which is known to be diamagnetic and superconducting only below T = 3.4 K and 0.03T.[5] Indium is used for soldering the gold wire as lead-tin solder dissolves the gold wire. Figure 2(c) shows a Hall bar sample mounted and wired on a header. Short pieces of 24-gauge wire were placed in the top of the header pins to facilitate soldering the gold wires. After plugging the sample header into the socket, a short piece plastic straw is wrapped around the bottom of the probe and fixed with Kapton tape so that the straw extends below the sample and protects the fragile sample and gold wires during insertion of the probe into the instrument. A standard plastic straw is cut to a length of ~ 5 cm and split lengthwise to accommodate the larger diameter of the probe end.

The probe can be inserted into the instrument after adjusting the position of the adjustable flange at the top end of the probe shown in Fig. 1(a). The sample should be located 128.0 ± 1.5 cm below the adjustable flange to position it in the center of the magnet, however, that distance should be checked as it may be different depending on the particular instrument used.

Transport measurements were made using the External Device Controller (EDC) plug-in in Quantum Design's Multiview operating system. In order to run the EDC plug-in, the PC computer was interfaced with a Keithley 2180 NanoVoltmeter through a GPIB interface. Multiview communicates with the voltmeter by making use of the EDC plug-in. A Delphi7 source code containing the GPIB communications commands was compiled to generate a ".dll" file. The .dll file was then called by the EDCInitialize and EDCExecute sequence commands in Multiview (see the Multiview manual).

### III. RESULTS AND DISCUSSION

Using this device we show results of magnetoresistance (MR) measurements on a thin film of the topological insulator $Bi_2Te_2Se$, where MR(B) = [R(B) − R(B=0)] / R(B=0). Weak antilocalisation (WAL) is expected to arise in those materials due to the destructive interference that occurs between oppositely traveling chiral electron wavefunctions on the surface of the film,[6] as well as due to strong spin-orbit coupling in the bulk[7]. This has been reported previously in $Bi_2Se_3$[4,8,9] and $Bi_2Te_3$[10]. MR measurements were performed at different temperatures in less than 10 hours for one sample. Results of the MR measurements as a function of applied magnetic fields and at various temperatures are shown below in Fig. 3.

The plots of MR(B) show a sharp WAL cusp at B = 0 developing as the temperature is lowered toward T = 2 K. This is potential evidence of the presence of surface Dirac electrons.[6] On the surface of a topological insulator the suppression of back scattering enhances the quantum corrections to the conductivity. Oppositely propagating helical electron channels interfere destructively thus making the electron wavefunction anti-localized leading to a smaller resistance.[6] The application of a perpendicular magnetic field destroys time-reversal symmetry allowing back-scattering to occur again. This is translated into a sharp increase in the resistance of the material when the magnetic field is increased.[11]

In order to fit the MR it is more convenient to plot the magnetoconductance, $\Delta G$, shown in Fig. 4. The magnetoconductance curves were then fit to the Hikami-Larkin-Nagaoka (HNL) model[12] for weak anti-localization:

$$\Delta G = \alpha \frac{e^2}{2\pi^2 \hbar} \left[ \psi \left( \frac{\hbar}{4eBL^2} + \frac{1}{2} \right) - \ln \left( \frac{\hbar}{4eBL^2} \right) \right],$$

where $\psi$ is a digamma function dominating the low field behavior of the magnetoconductance, B is the magnetic field, $L$ is the phase coherence length, and $\alpha$ is a constant related to the Berry phase of the electron wavefunction. Each surface is expected to give an $\alpha \sim$ -0.5 contribution, resulting in a total of $\alpha \sim$ -1 from both surfaces.[6]

The HLN model was fit to the magnetoconductance curves measured below T =30 K in Fig. 4. The fits yielded $\alpha \sim$ -0.2. This implies a non-zero Berry phase contribution from one surface. This discrepancy could be due to bulk spin-orbit interaction terms that were not taken into account.[7] Another explanation can be sought in the fact that surface to bulk scattering can result in surface channels being coupled to each other as well as to bulk channels.[9] The behavior of the phase coherence length $L$ was studied versus temperature, (shown in the inset of Fig. 4). It was observed that $L$ decays proportionally to $T^{-0.44\pm0.07}$ suggesting evidence of 2D transport[13]. According to ref. 13, in a 2D system, $L$ should decay proportionally to $T^{-0.5}$ as a result of electron-electron collisions occurring at low temperatures. In total, these measurements indicate the presence of 2D transport and a non-zero Berry phase that is evidence of the presence of Dirac surface electrons.

## IV. CONCLUDING REMARKS

In conclusion, Quantum Design's MPMS magnetometer optional transport probe can be modified to offer users more flexibility. This allows one to perform electrical measurements under well-controlled temperature conditions in short periods of time. Of additional importance is the fact that this allows one to perform magnetic and transport characterization in a single chamber, hence in the same environment. This is of great use for experiments that seek to explore electron transport that is correlated to magnetism. We demonstrated the utility of this device by characterizing the transport of a $Bi_2Te_2Se$ thin film. The observed behavior was analyzed and revealed surface contributions to the magnetoresistance.

## ACKN0WLEDGMENTS

This work was supported by a grant from the National Science Foundation DMR-0907007. PW and FK were supported in part by the MRSEC Program of the National Science Foundation under award number DMR – 0819762. JSM is partly supported by NSF (DMR 0504158) and ONR (N00014-09-1-0177) grants. We thank J. O'Brien of Quantum Design, Inc. for useful conversations and T. Hussey for technical support. This work is dedicated to the late Si Foner who was awarded the Joseph F. Keithley Award for inventing the modern vibrating sample magnetometer (VSM) and who was an editor of the Review of Scientific Instruments. The small diameter socket-header device was originally designed by Foner and one of us (DH) as a means of measuring magnetotransport in Foner's 68 T pulsed-field magnet at the MIT Francis Bitter National Magnet Lab.[14]

**REFERENCES**

[1] H. Ohno, A. Shen, F. Matsukura, A. Oiwa, A. Endo, S. Katsumoto, and Y. Iye, Appl. Phys. Lett. **69**, 363 (1996).

[2] T.J. Nummy, S.P. Bennett, T. Cardinal, and D. Heiman, Appl. Phys. Lett. **99**, 252506 (2011).

[3] Y. Zhang, Y.Tan, H. L. Stormer, and P. Kim, Nature **438**, 201 (2005).

[4] J.G. Checklsky, Y.S. Hor, M.H. Liu, D.X. Qu, R.J. Cava, and N.P. Ong, Phys. Rev. Lett. **103**, 246604 (2009).

[5] N.W. Ashcroft and N.D. Mermin, *Solid State Physics*, 1st Edition (Harcourt College Publishers, New York, 1976), p. 729.

[6] H.Z. Lu, J. Shi, S.Q. Shen, Phys. Rev. Lett. **107**, 076801 (2011)

[7] G. Bergmann, Solid State Comm. **42**, 815 (1982).

[8] J. Chen, H.J. Qin, F. Yang, J. Liu, T. Guan, F.M. Qu, G.H. Zhang, J.R. Shi, X.C. Xie, C.L. Yang, K.H. Wu, Y.Q. Li, and L. Lu, Phys. Rev. Lett. **105**, 176602 (2010).

[9] H. Steinberg, J.B. Laloë, V. Fatemi,  J.S. Moodera , and P. Jarillo-Herrero, Phys. Rev. B **84**, 233101 (2011).

[10] H.T. He, G. Wang, T. Zhang, I.K. Sou, G.K. Wong, J.N. Wang, H.Z. Lu, S.Q. Shen, and F.C. Zhang, Phys. Rev. Lett. **106,** 166805 (2011).

[11] F.E. Meijer, A.F. Morpurgo, T.M. Klapwijk, J. Nitta, Phys. Rev. Lett. **94**, 186805 (2005).

[12] S. Hikami, A.I. Larkin, and Y. Nagaoka, Prog. Theor. Phys. **63**, 707 (1980).

[13] B.L. Altshuler, A.G. Aronov, and D.E. Khmelnitsky, J. Phys. C: Solid State Phys. **15**, 7367 (1982).

[14] D. Heiman and S. Foner (unpublished).

**FIGURES**

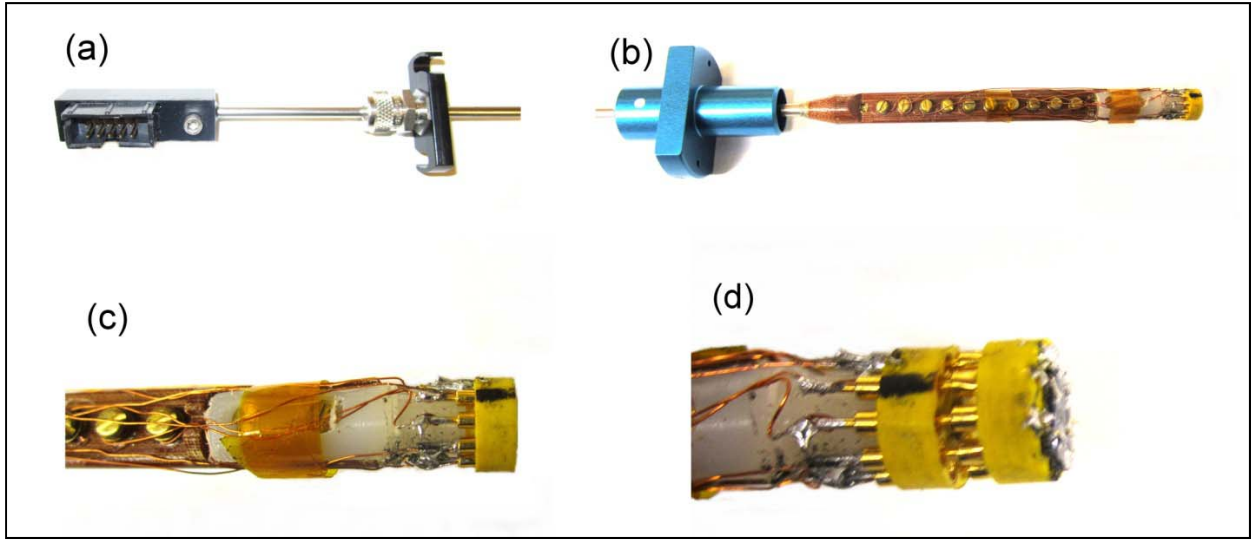

FIG. 1. Pictures of the electrical transport probe modified for use with a Quantum Design MPMS SQUID magnetometer. (a) Top room-temperature end of the probe showing the rectangular 10-pin socket for attaching an electrical cable leading to measurement electronics. (b) Bottom of probe showing the 10 small screws for electrical connections. (c) Bottom end of probe showing the circular 10-pin Augat header attached and wired up as a socket. (d) Close up of socket with header and attached sample plugged in.

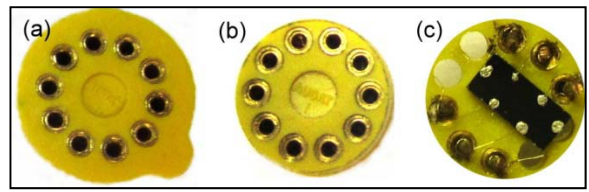

FIG. 2. Top views of circular 10-pin Augat headers. (a) Header before machining the outside diameter. (b) Header after machining the outside diameter to a dimension of about 7.5 - 8.0 mm diameter. (c) Machined header with Hall-bar sample mounted and wired. Four pins have been removed to eliminate overlapping of the sample on the unused pins.

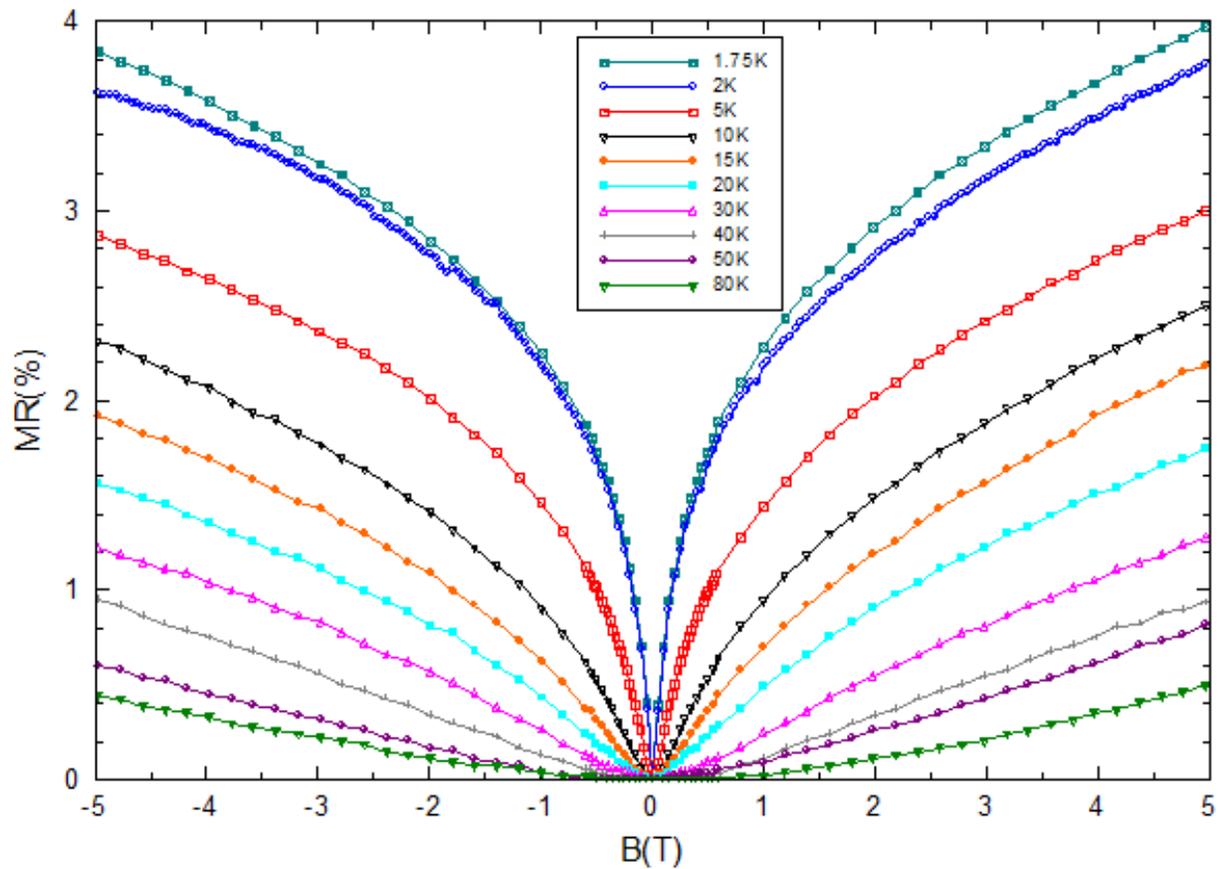

FIG. 3. (Color Online). Magnetoresistance of 15 nm thick $Bi_2Te_2Se$ film versus magnetic field applied to perpendicular to film layer at temperatures between T = 1.75 K and 80 K.

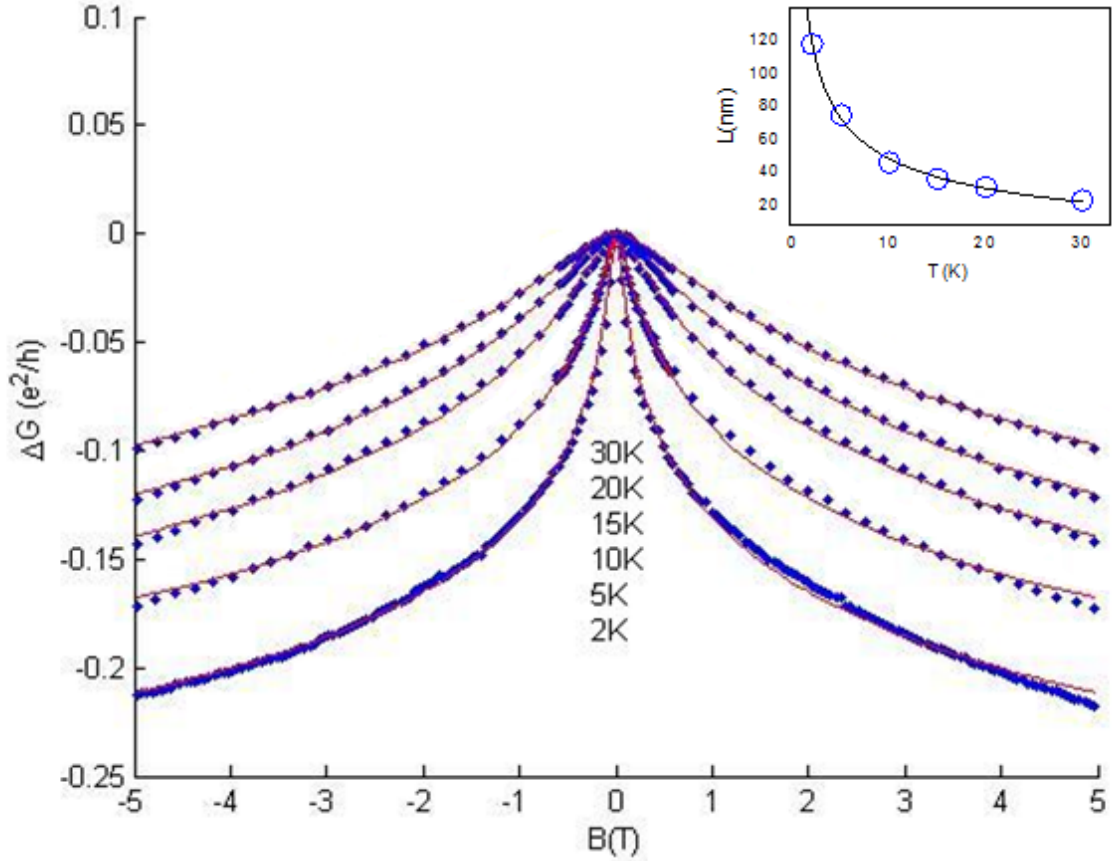

FIG. 4. Magnetoconductance of 15 nm thick $Bi_2Te_2Se$ film versus magnetic field applied to perpendicular to film layer at temperatures between T = 2 K and 30 K. The points are experimental data and the solid curves are fits to the weak anti-localization model described in the text. The inset plots the phase coherence length (L) versus temperature obtained by fitting the magnetoconductance data. The solid curve is a fit of the L(T) data to the function $T^{-\beta}$ where $\beta = 0.44 \pm 0.07$.